\documentclass[]{spie}  

\usepackage{amsmath,amsfonts,amssymb}
\usepackage{graphicx}
\usepackage{xspace}
\usepackage[colorlinks=true, allcolors=blue]{hyperref}
\usepackage[sort,super]{natbib}
\usepackage{eurosym}

\newcommand{\uas}{\ensuremath{\mu\mbox{as}}\xspace}

\newcommand{\gaia}{\textit{Gaia}\xspace}

\newcommand{\theia}{\textit{Theia}\xspace}

\newcommand{\hip}{\textit{Hipparcos}\xspace}

\newcommand{\hst}{\textit{HST}}

\newcommand{\simMission}{\textit{SIM}}

\newcommand{\euclid}{\textit {Euclid}}

\newcommand{\jwst}{\textit{JWST}} 
\newcommand{\wfirst}{\textit{Roman}}
\newcommand{\rubin}{\textit{Rubin}}

\title{Challenges in focal plane and telescope calibration for High-Precision Space Astrometry}

\author[a]{Fabien Malbet}
\author[a]{Manon Lizzana}
\author[a]{Fabrice Pancher}
\author[a]{Sébastien Soler}
\affil[a]{Univ.\ Grenoble Alpes, CNRS, IPAG, 38000 Grenoble, France}
\author[b]{Alain Léger}
\affil[b]{Univ.\ Paris-Saclay, CNRS, Institut d'astrophysique spatiale, Orsay, France}
\author[c]{Thierry Lépine}
\affil[c]{Institut d'Optique \& Hubert
  Curien Lab, Univ.\ de Lyon, Saint-Etienne, France}
\author[d]{Gary A.\ Mamon}
\affil[d]{Sorbonne Université, CNRS, Institut d’Astrophysique de Paris, Paris, France}
\author[e]{Alessandro Sozzetti}
\affil[e]{Obs.\ Torino/INAF, Pino Torinese, Italy}
\author[e]{Alberto Riva}
\author[e]{Deborah Busonero}
\author[f]{Lucas Labadie}
\affil[f]{Univ.\ of  Cologne, Cologne, Germany}
\author[g]{Pierre-Olivier Lagage}
\affil[g]{Univ. Paris-Saclay, CEA, Saclay, France}
\author[h]{Renaud Goullioud}
\affil[h]{Jet Propulsion Laboratory, California Institute of  Technology, Pasadena, CA, USA}

\authorinfo{Send correspondence to FM using the email address
  \texttt{Fabien.Malbet@univ-grenoble-alpes.fr}}

\begin{document} 
\maketitle

\begin{abstract}
  With sub-microarcsecond angular accuracy, the \theia telescope will
  be capable of revealing the architectures of nearby exoplanetary
  systems down to the mass of Earth. This research addresses the
  challenges inherent in space astrometry missions, focusing on focal
  plane calibration and telescope optical distortion. We propose to
  assess the future feasibility of large-format detectors (50 to 200
  megapixels) in a controlled laboratory environment. The aim is to
  improve the architecture of the focal plane while ensuring that
  specifications are met. The use of field stars as metrological
  sources for calibrating the optical distortion of the field may help
  to constrain telescope stability. The paper concludes with an
  attempt to confirm in the laboratory the performance predicted by
  simulations. We will also address the possibility of using such
  techniques with a dedicated instrument for the Habitable World
  Observatory.
\end{abstract}

\keywords{astronomy, astrophysics, dark matter, exoplanet, astrometry,
  differential, visible, high precision, space mission}

\section{INTRODUCTION}
\label{sec:intro}  

ESA's \hip and \gaia global astrometry scanning missions have
revolutionized our understanding of the Solar Neighborhood and the
Milky Way, providing fundamental new foundations for numerous fields of
astronomy.

Global astrometry involves connecting stars at large angular distances
in a network where each star is linked to several others in all
directions. The network closure condition ensures the reduction of
positional errors for all stars\citep{1980CeMec..22..153K}. The \hip
mission\cite{1997ESASP1200.....E} demonstrated its capability to
achieve global measurements, such as positions and changes in position
due to proper motion and parallax, determined within a reference
system consistently defined across the entire sky for a large number of
objects. Milliarcsecond precision was achieved using a continuously
scanning satellite that observes two directions simultaneously. 

Similar to \hip but with $100$ times greater accuracy, $1,000$ times
greater limiting magnitude and $10,000$ times greater number of stars
observed, \gaia \citep{2016A&A...595A...1G} consists of two telescopes
offering two directions of observation with a fixed wide angle between
them. The spacecraft rotates continuously around an axis perpendicular
to the lines of sight of the two telescopes. This axis of rotation
slightly precesses across the sky while maintaining the same angle to
the Sun. By precisely measuring the relative positions of objects in
both directions, a rigid reference system is obtained. The two main
features of the telescope are: a $1.45\times0.5$,m primary mirror for
each telescope and a $1.0 \times 0.5$,m focal plane onto which the
light from both telescopes is projected. This array comprises over 900
megapixels. On average, each celestial object has been observed about
70 times over the five-year nominal mission, which has been extended
to ten years, doubling the number of observations.

For specific astronomical objects, the accuracy of astrometric
measurements can be enhanced by dedicating more observing time to
determining their relative position to the stars in the field of view,
known as relative astrometry or differential astrometry.

The topic of high-precision astrometry emerged from the \emph{Space
  Interferometry Mission (SIM)} in the late 2000s, aimed at detecting
Earth-like exoplanets while maintaining the methodology to achieve
absolute astrometry. Following the cancellation of SIM by NASA in
2010, a small team proposed a new concept to address high-precision
astrometry called the \emph{Nearby Earth Astrometry Telescope
  (NEAT)}. NEAT, consisting of a single off-axis mirror directing
light onto a detector in a formation flight
configuration\cite{Malbet+12}, was presented for the M3 mission call
at ESA. Subsequently, the \theia concept, featuring a single
spacecraft with a Korsch three-mirror anastigmatic (TMA) telescope, a
single focal plane, and instrument metrology subsystems, was studied
and submitted, albeit unsuccessfully, to the M4, M5, and M7 medium
mission calls at ESA\cite{2017arXiv170701348T}.

This article briefly reviews the scientific arguments in favor of
high-precision space astrometry for exoplanets, then explores in more
detail the challenges posed by focal plane and telescope
calibration. The case for dark matter has been adressed in previous
works\cite{2022SPIE12180E..1FM, Malbet+21} and will not be discussed
hereafter.

\section{Exploring Earth-like Planets Orbiting Nearby Sun-like Stars}

\begin{figure}[t]
  \centering
  \includegraphics[width=0.75\hsize]{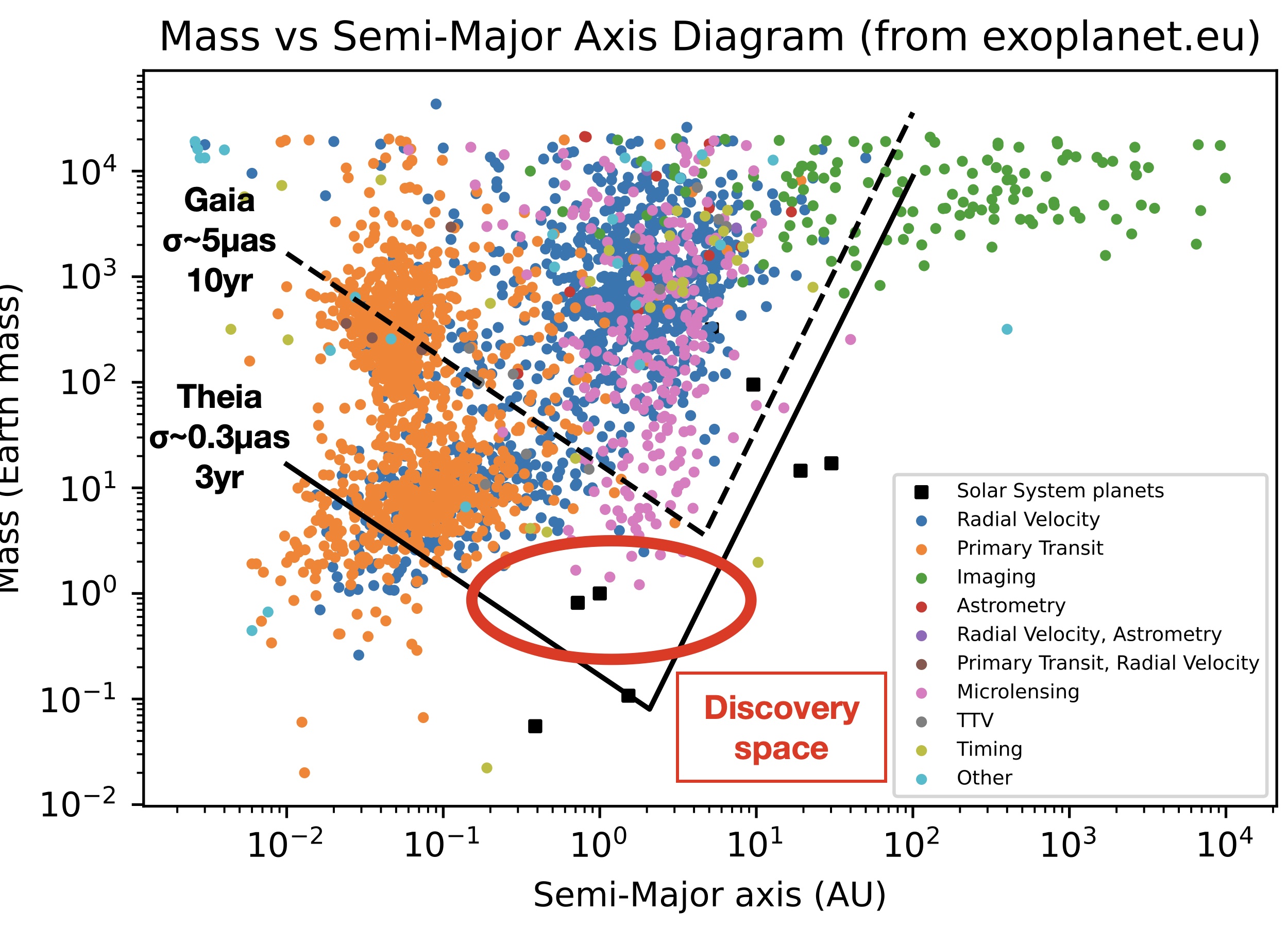}
  \caption{Exoplanets discovery space.}
  \label{fig:exoplanets-param-space}
\end{figure}

The detection and atmospheric characterization of temperate,
potentially habitable terrestrial planets orbiting the nearest
Sun-like stars is a key science theme in both the final
recommendations of the senior Scientific Committee for ESA’s long-term
scientific plan Voyage 2050 and the US Decadal Survey on Astronomy and
Astrophysics 2020 \citep[(Astro2020)]{2021pdaa.book.....N}. \theia's
exceptional single-measurement positional precision in pointed,
differential astrometric mode ($< 1\,\mu$as) will enable the detection
and high-confidence ($\geq 3\sigma$) true mass determination of Earths
and Super-Earths (M = 1--5 M$_\oplus$) in the Habitable Zone (HZ) of
the $\sim60$ nearest solar-type stars
(Fig.~\ref{fig:exoplanets-param-space}). This will be achieved through
high-cadence observations ($\approx 100$ visits over 3 years) of each target and
$\geq 100$ reference stars\cite{Malbet+21} (see Sect.~\ref{sec:dist-calibr-exper}).

Ground-based extreme-precision, sub-m/s Doppler techniques are
expected to provide a global census of temperate terrestrial planets
around nearby late-type M dwarfs in the next years. However, for
solar-type stars, the Doppler method might be limited by stellar
activity, potentially missing any HZ Earth-mass companions whose
orbits are not close to edge-on. \theia's astrometric sensitivity will
allow us to achieve three critical goals in exoplanetary science:
\begin{enumerate}
\item \theia astrometry will enable the determination of {\bf the true
    mass function of temperate 1--5 M$_\oplus$ rocky planets around
    solar-type stars}, {\em which is currently unknown}. 
\item By measuring the true masses and full three-dimensional
  architecture in multiple systems, \theia will facilitate the study
  of {\bf the full demographics of planetary systems that host
    temperate terrestrial planets around the nearest Sun-like stars},
  in high synergy with \gaia and Doppler surveys. 
\item The temperate terrestrial planets detected by \theia will
  provide {\bf the fundamental input target list for direct-imaging
    and spectroscopic missions aimed at searching for atmospheric
    biomarkers}. 
\end{enumerate}

For such ambitious space observatories, either in the
optical/near-infrared (e.g., NASA's proposed flagship missions HabEx,
LUVOIR) or in the thermal infrared (e.g., the LIFE concept for an
ESA's L-class mission), providing their target list will be
crucial. This will avoid their research phase (about 50\% of their
mission time) and allow them to invest all their precious observing
time performing spectroscopy of atmospheres, knowing exactly where to
look. Prior knowledge of the true masses will also be essential in
interpreting any molecular detections in these atmospheres. 

\begin{figure}[t]
  \centering
  \includegraphics[width=0.45\hsize]{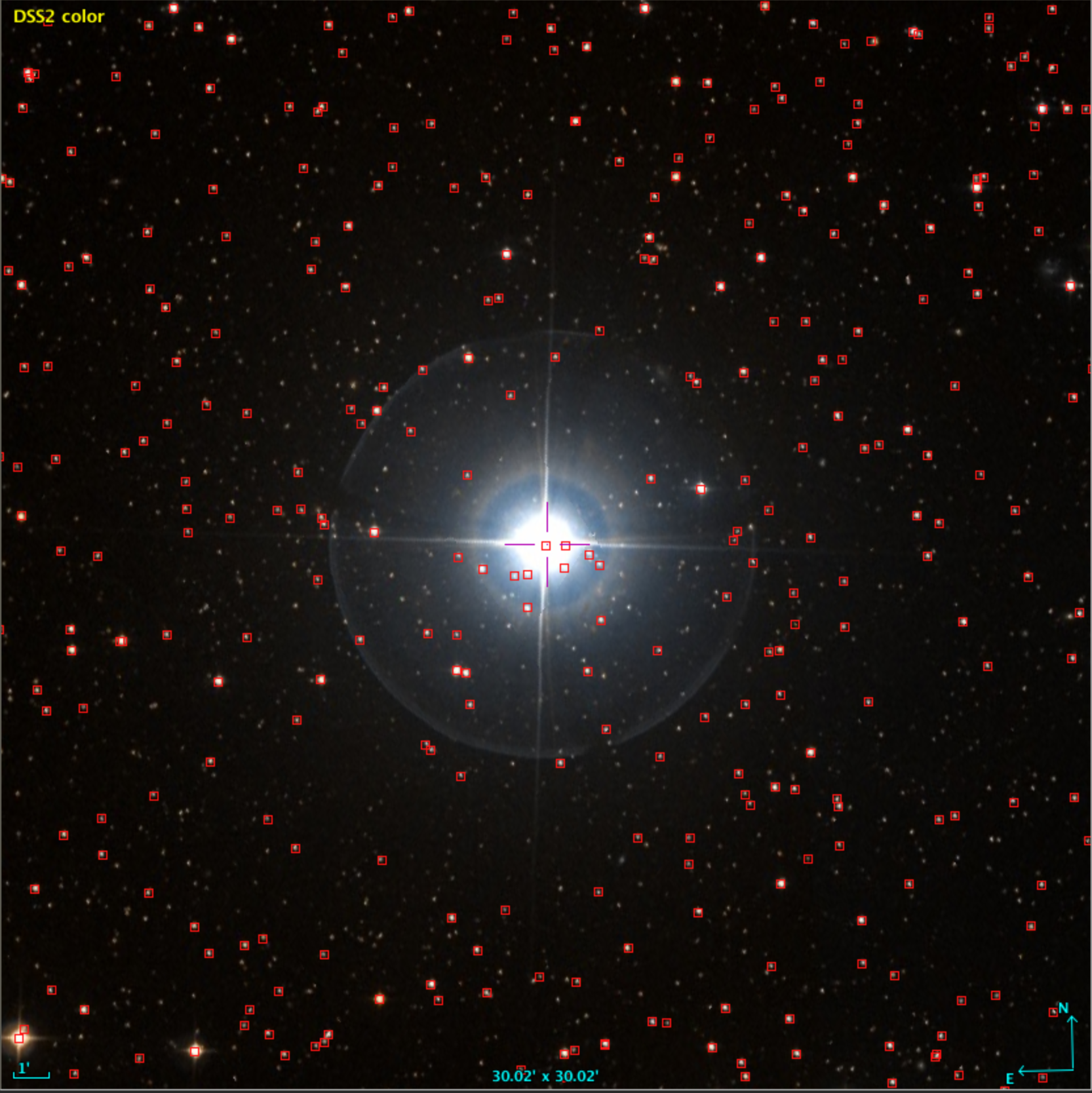}
  \label{fig:ups-and}
  \bigskip
  \caption{Image from Upsilon Andromedae from Digital Sky Survey. The field of view is $30' \times
30'$ and the red boxes correspond to GAIA DR3 sources. 261 sources are
brighter than $G=14\,\mathrm{mag}$ within the  30'\,$\times$\,30'
field of view.}
\end{figure}

Estimates of \theia's sensitivity\cite{Meunier&Lagrange22}
indicate that the median detectable mass across the full HZ for the
\theia\ stellar sample is $\simeq 1.1$ M$\oplus$. If current
extrapolations of the occurrence rate of true Earth-like
planets\cite{Bryson+21} at $37_{-21}^{+48}\,\%$ are accurate, we expect
to detect between 9 and 57 such planets. Furthermore, the number of HZ
Earths per star could be greater than one, as a typical HZ can
dynamically sustain more than one planet.

\theia is a single field, visible-wavelength ($400-900$ nm)
differential astrometry mission, meaning that the derived astrometric
parameters for the target stars in a field will have position,
parallax and proper motion relative to a local reference frame tied to
a global one. At the time of the \theia mission, the most accurate and
complete optical reference frame will be that of the \gaia catalog. By
using \gaia global astrometry parameters as priors, the astrometric
solution of all the stars observed by \theia will be automatically
tied to the \gaia frame, without the need of forcing physical priors
on sources such as quasars or remote giant stars.  

These characteristics make \theia much superior to several
competitors. The deep fields will achieve 23 times the proper motion
precision of \gaia (10
years)\footnote{https://www.cosmos.esa.int/web/gaia/science-performance\#astrometric\%20performance}
and 14 times that of \hst\cite{Vitral+22} (10 years). In the
bright-star regime\cite{Malbet+21}, \theia's 1-$\mu$as precision in
1-hr integration at the reference value $R=10$ mag exceeds that of
\gaia\cite{Lindegren+2018}, \wfirst\cite{sanderson+2019}, and {\em
  VLTI/GRAVITY}\cite{Gravity+2021} by factors of $\sim40-50$,
$\sim10-20$, and $\sim30-100$ respectively, and other
missions/instruments (e.g., \hst, \jwst, \emph{ELT/MICADO}, \rubin) by
even larger factors.

\section{\theia main mission characteristics}
\label{sec:mission}

The baseline \theia Payload Module leverages the heritage knowledge of
consortium members in space mission concepts such as \gaia, \hst/FGS,
\simMission, NEAT/M3, \theia/M4+M5, and \euclid. The \euclid-like
mission, featuring a Korsch three-mirror anastigmatic (TMA) telescope,
a single focal plane, and instrument metrology
subsystems\cite{Malbet+21}, is favored to avoid formation flying
constraints. The optimal configuration is a 0.8-m on-axis TMA
telescope operating at visible wavelengths, as described in the
\theia/M5 proposal\cite{2017arXiv170701348T}, but with an upgraded
optical design. The optics remain coaxial, but the field of view
center is shifted by 0.45\,deg to ensure the light beam avoids the
plane mirror after reflection on M3. 

Compared to the proposed \theia/M5 mission concept, a significant
advancement is the incorporation of new types of CMOS (complementary
metal oxide semiconductor) detectors, which
allow up to $10^9$ small-size ($\sim 4\mu$m) pixels with
well-controlled systematics. These detectors can read pre-determined
windows around objects with pixel readout rates $\geq$1,kHz to prevent
saturation of bright stars. Such detectors simplify the payload
considerably by requiring fewer or even a single detector and read-out
electronics to cover the necessary $\sim0.5^\circ$ field-of-view (FoV)
in the focal plane array (FPA). They also shorten the focal length to
approximately $13\,\rm m$.

Achieving sub-\uas-level differential astrometric precision requires
controlling all factors that affect the relative positions of the
Nyquist-sampled point-spread function (with an apparent size of
136\,mas for a 0.8-m telescope in the visible). The precision of
relative position determination on the detector depends on photon
noise (limited by the reference stars), the geometric stability of the
focal plane array, optical aberrations, and variations in detector
response between pixels. This necessitates a fundamental precision
requirement of approximately $5.10^{-6}$ pixel. 

To monitor the various sources of distortion in the FPA and correct
the associated systematic errors, \theia will rely on metrology
powered by a laser source through optical fibers placed behind the
nearest mirror and projecting Young's fringes onto the detector(s) that
can be used to measure the relative positions of each pixel. 

In addition the telescope geometry is expected to vary, even at very
stable environments such as L2 and therefore a baseline telescope
metrology subsystem was proposed, based on a concept of linear displacement
interferometers and piezo activator applied independently on each linear
element of the structure. However, a new method to derive
the astrometry signal using the reference stars allows us to measure
the telescope distortion described in Sect.~\ref{sec:diffastrometry}
and therefore considerably relax the initial requirement set for the M5
proposition from several hours to fraction of a second.

\section{High precision differential astrometry for exoplanet detection}
\label{sec:diffastrometry}

Table \ref{tab:mission-parameters} gives the main parameters required
to perform the detection of an Earth-like planet in system located at
10\,pc from the Sun with \theia and HWO. The difference between the
two telescope is the size of the telescope and therefore the size of
the diffraction-limited point-spread function. The factor 10 between
the two means that we can go deeper with HWO to look for faint
reference stars in a limited field of view, but also the required
precision on the focal plane is not as strong as with \theia ($\sim
5.10^{-5}$ pixels instead of $\sim 5.10^{-6}$ pixels.

\begin{table}
  \centering
    \caption{Focal plane parameters relevant to the different space
      missions}
    \label{tab:mission-parameters}
    \smallskip
  \begin{tabular}{lll}
    \hline
     &Theia &Habitable World Observatory (HWO)\\
    \hline     \hline
    Wavelength of operation
     & $400\,\mathrm{nm} \le \lambda \le 900\,\mathrm{nm}$
     &UV, visible, and infrared\\*[1ex]
    
    Mission status
     & Proposed project (next call M8)
     &NASA flagship mission, \\
     &&foressen launch $\sim2040$ \\*[1ex]

    Telescope diameter
     & 0.8\,m
     & between 6.5\,m and 8\,m\\*[1ex]

    Diffraction limit
     & 134\,mas
     &14\,mas \\*[1ex]

    Signature of an Exo-Earth
     & $\sim 5.10^{-6}$ pixels
     & $\sim 5.10^{-5}$ pixels\\
    @ 10\, pc ($0.3\,\uas$)\\*[1ex]

    Field of view
     & $0.5^{\rm o} \times 0.5^{\rm o}$ or 30'\,$\times$\,30'
     & 2'\,$\times$\,3' (wide field imager, TBC)\\
     & $25.6\,\mathrm{kpx} \times 25.6\,\mathrm{kpx}$
     & $17.1\,\mathrm{kpx} \times 25.6\,\mathrm{kpx}$ \\*[1ex]

    Faintest reference stars with
     & $m_V(\mathrm{ref}) \le 14$
     & $m_V(\mathrm{ref}) \le 20$ \\
    a $m_V=6$ science target\\*[1ex]

    Resulting required minimum 
     &$10^3$ & $10^5$\\
    dynamic range&&\\
    \hline
  \end{tabular}
\end{table}

In the case of the search for Earth-like planets orbiting nearby
Sun-like stars, one can take the example of $\upsilon$ Andromedae
(Fig~\ref{fig:ups-and}) where we see the field of view of $30' \times
30'$ of the Digital Shy Survey image around this target. Using \gaia
DR3 catalog, we can retrieve 261 sources fainter than
$G_\mathrm{max} = 14$. 

\section{Laboratory testbeds at IPAG}

In order to increase the Technology Readyness Level (TRL) of the
mission, we have decided to implement several laboratory testbeds at
IPAG with the help of the LabEx FOCUS and CNES.

\subsection{Detector classical characterization}

\begin{figure}
  \centering
  \includegraphics[width = 0.6\hsize]{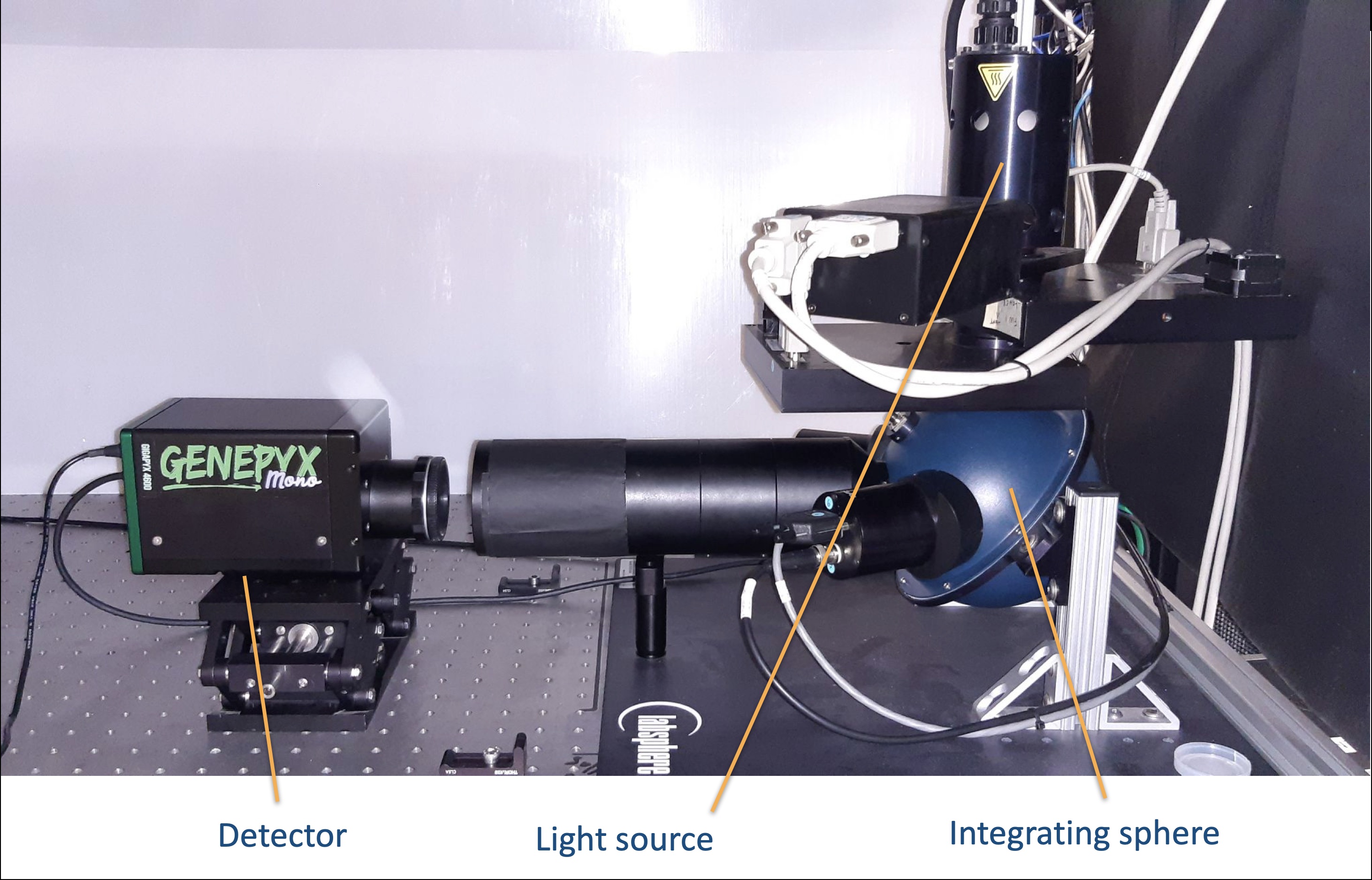}
  \includegraphics[width = 0.95\hsize]{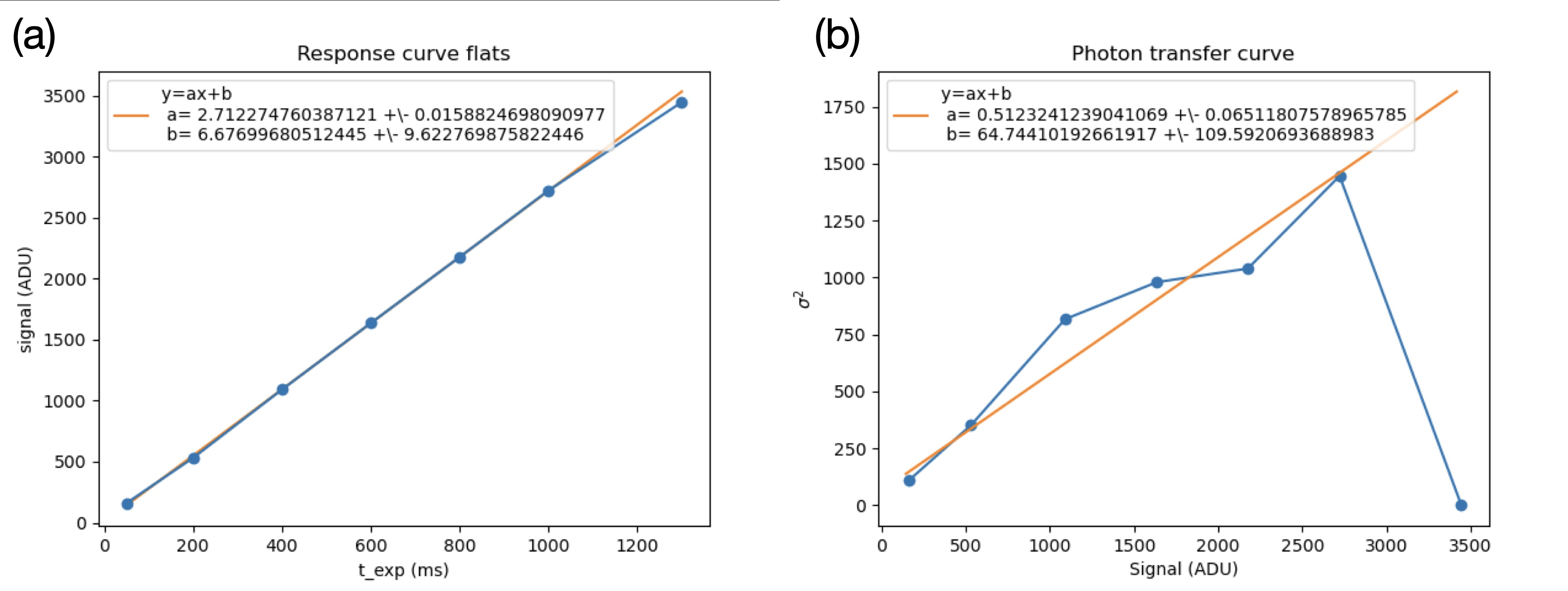}
  \caption{Top panel : Detector characterization testbed with on the
    left part the Gigapyx detector embedded in the GenePyx camera and on
    the right part the integrating sphere. Bottom figures : (a) the
    response curve abtained with images of flats; (b) photon transfer curve.}
  \label{fig:detector-classical-calib}
\end{figure}

We have procured a 46 megapixel detector from Pyxalis (Gigapyx 4600)
in order to test the new type of large CMOS detector and also redo the
interferometric measurements made by Crouzier et
al.\cite{Crouzier+16}.

We have started a global characterization of this detector
(Fig.~\ref{fig:detector-classical-calib}) : linearity, dark current,
readout noise, gain, map of defective pixels, etc.

\subsection{Detector pixel geometry using a interferometric
  calibration}

In a real detector the stars are misaligned because of fine
pixel structure, quantum efficiency local variations, etc. This is
why, we need to calibrate them using the  following sequence :
\begin{itemize}
\item Fibres create Young's fringes on the detector ;
\item Phase modulators make the fringes scroll ;
\item The modulation observed by each pixel 
  provides the position of the centroids 
\item The precision required is below 5e-6 px (corresponding to $0.3\,\uas$, an exo-Earth signal)
\end{itemize}

The proof of concept on small matrices has been performed in
2016 by Crouzier et al.\cite{Crouzier+16} but the performance needs to be validatednow on
large matrices. This is why we have developed a simplified testbed at
IPAG using the components from Crouzier.

\subsection{Distortion calibration experiment}
\label{sec:dist-calibr-exper}

The optical components of the telescope induce optical aberrations that can
shift the positions of stars on the detector of thousand of pixels
\cite{Malbet+12}. Whereas the required precision is
$0.3\,\uas$ corresponding to $5.10^{-6}~\mathrm{px}$ on the detector, hence
the optical aberrations must be calibrated.

\begin{figure}
  \centering
  \fbox{\includegraphics[width=0.65\linewidth]{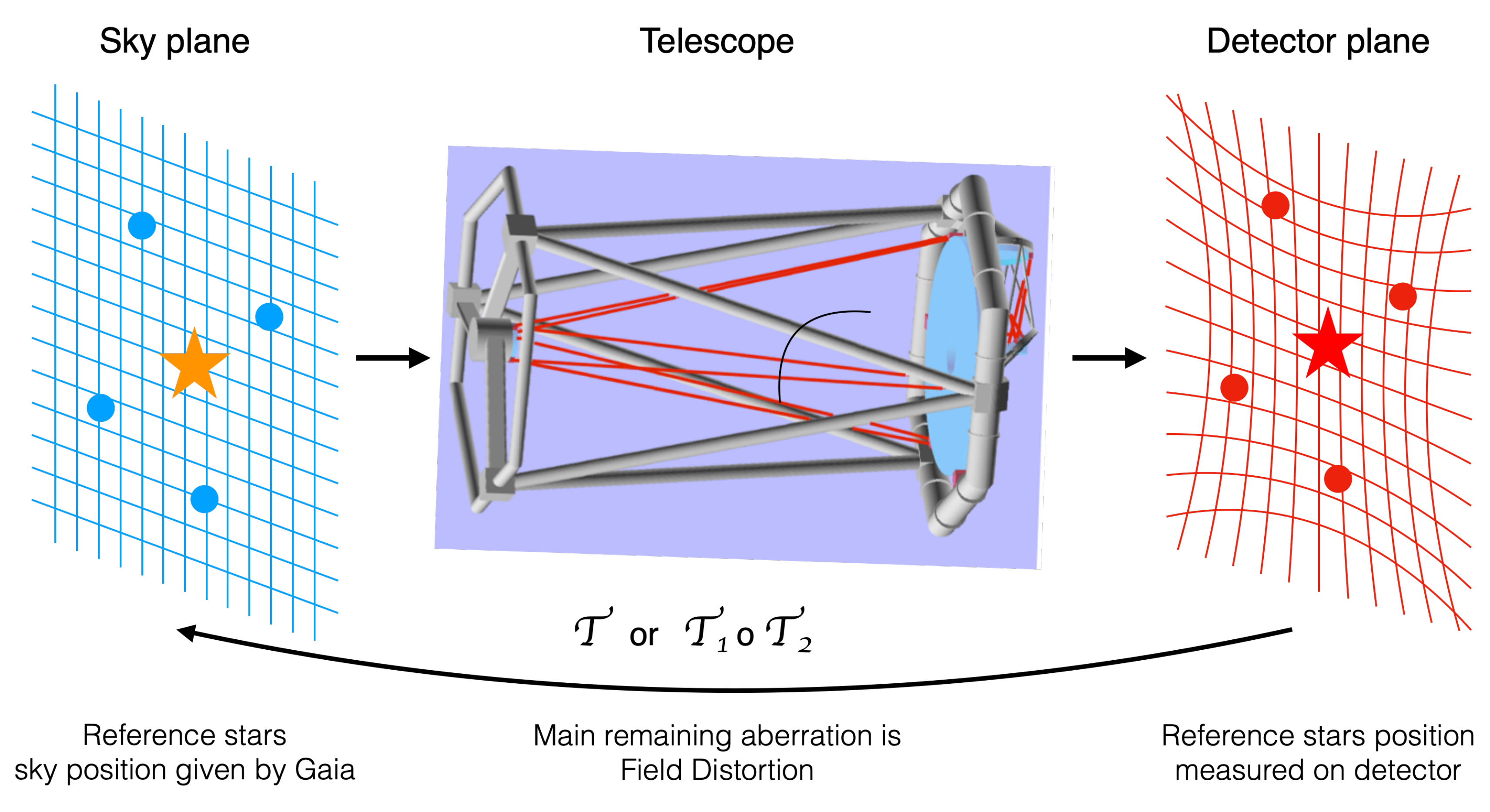}}
  \bigskip
  \caption{General illustration of the calibration of the
    distortion. The reference stars are used to adjust a bivariate
    polynom $\mathcal{T}$ (or a combination of two polynoms $\mathcal{T}_1$ and
    $\mathcal{T}_2$) that relates the positions of the detector plane to the
    sky positions. The motion of the target is observed using this
    transformation. }
  \label{fig:distorted_grid}
\end{figure}
  
Fixing this issue thanks to laser metrology is possible but add a lot
difficulties is the use the stars in the field of view (reference
stars) as metrology sources in order to compute the distortion
function. The distortion function links the positions of stars
on the sky with the positions of stars images on the
detector. We use bivariate polynomials to model the optical aberration
based on the knowledge of the reference star position by \gaia (see
Fig.~\ref{fig:distorted_grid}).

\begin{figure*}
  \centering
  \includegraphics[width=0.9\textwidth]{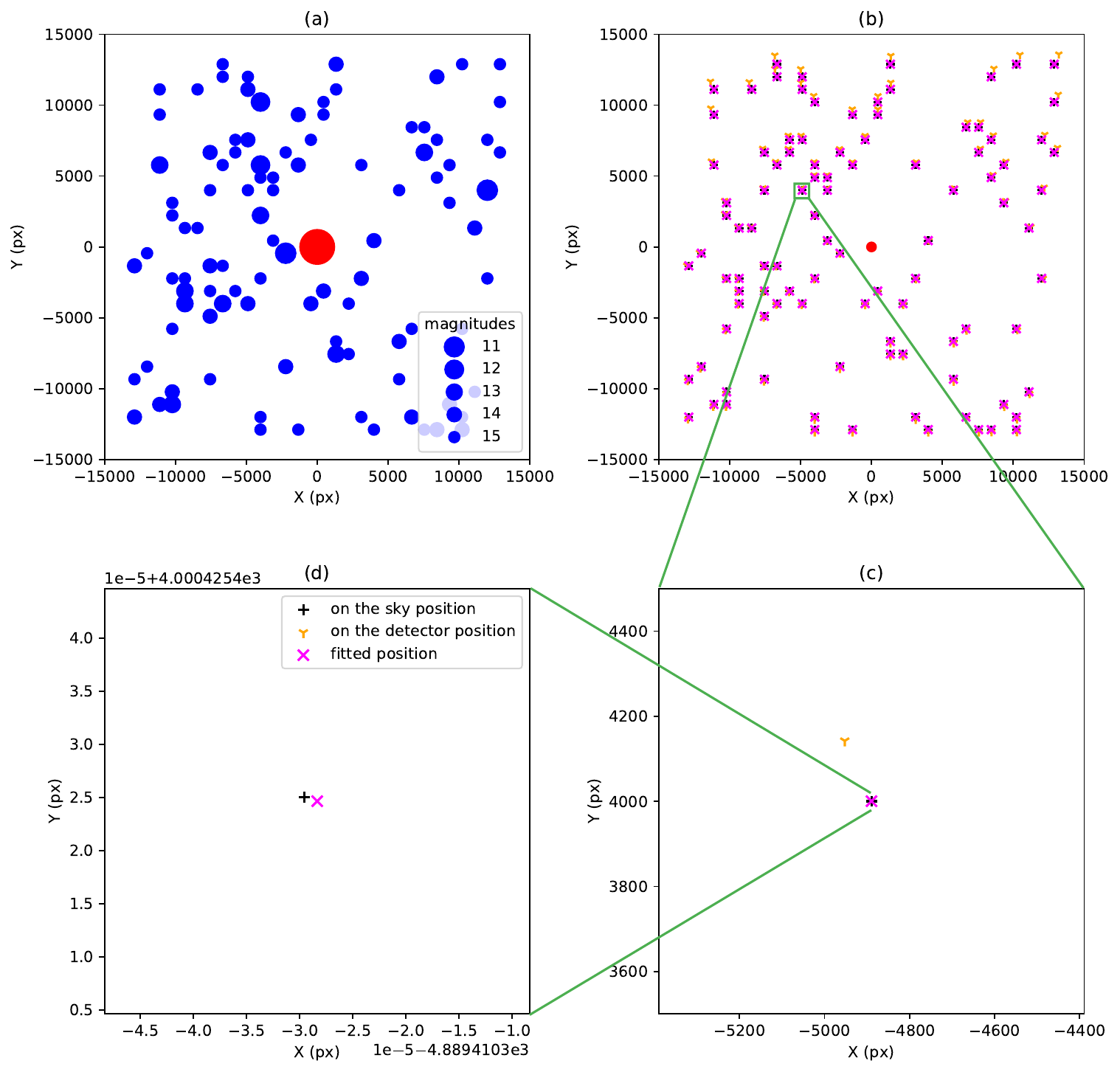}
  \caption{Example of on-sky and simulated fields at different
    magnification rate : (a) Sky field : the image of a star is a
    circle with size function of its magnitude; the red star is the
    target star; (b) On-sky and detector positions at full scale:
    detector positions are slightly shifted compared to on-sky
    positions due to distortion, but on-sky and fitted positions are
    indistinguishable; (c) at the scale of the distortion which reach
    a few hundred pixels, the on-sky and fitted positions are still
    indistinguishable; (d) at the micropixel scale the on-sky and
    fitted positions are separated. }
  \label{fig:field}
\end{figure*}

\begin{figure*}
    \centering
    \includegraphics[width=0.95\textwidth]{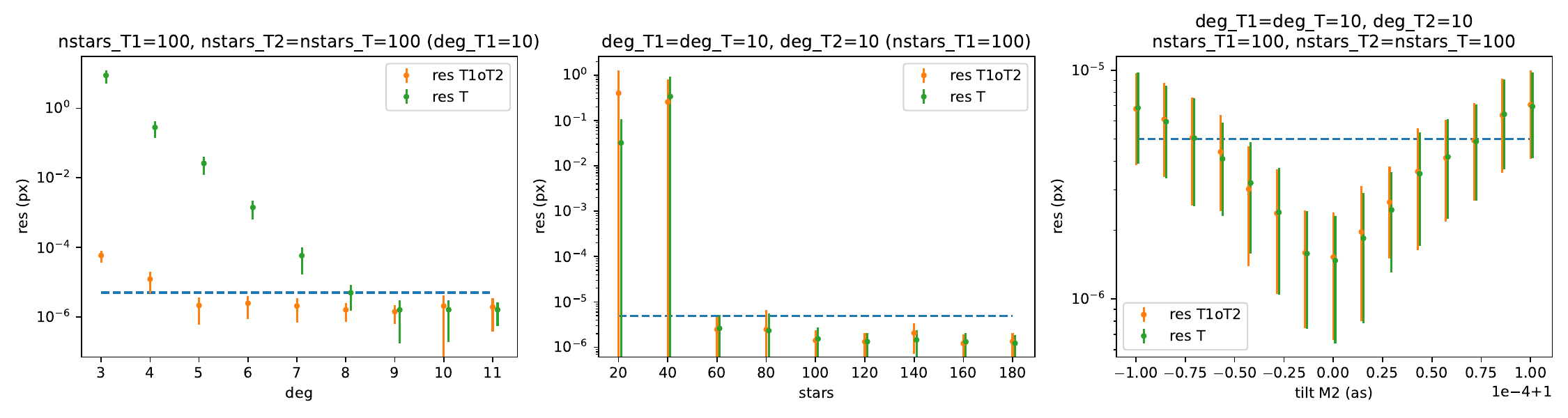}
    \caption{Evolution of the position residuals (i.e.\ the difference
      between the sky positions and the positions computed with the
      polynoms after a chi-squared adjustment on the grid of optical
      rays) with respect to the degree of the polynomial, the number
      of reference stars and the tilt of the M2 (expressed in arcsec). \texttt{n\_stars\_T},
      \texttt{n\_stars\_T1}, \texttt{n\_stars\_T2}, is the number of stars
      randomly chosen for the computation of the coefficients of the
      polynom of \texttt{deg} degree to perform the fit. See text for
      details. The dashed blue line corresponds to the specification
      for exo-Earth detection, i.e.\ $5.10^{-6}$ pixels.}
    \label{fig:residuals}
  \end{figure*}

The main results are presented in Fig.\,\ref{fig:residuals}. In our
simulations a precision of $5\,10^{-6}\,\mathrm{px}$ can be reached with about
100 stars and an 8$^{\text{th}}$ order polynomial, if the M2 tilt is below
$7\,10^{-5}$\, arcsec for that exposure. In addition, the residuals are almost
constant all over the field of view. The fit with $\mathcal{T}$ and $\mathcal{T}_1 \circ
\mathcal{T}_2$ seems to reach the same precision.

We can conclude that the stars in the field of view can be used as
metrology sources, and a 2D polynomial model seems to be able to
calibrate the telescope's distortion with a precision better than
$5.10^{-6}\,\mathrm{px}$.

The positions of the reference stars on the sky are
known with a certain error thanks to the \gaia catalog and the
positions on the detector will be measured by
the telescope as the photocenter of the diffraction spot. The
distortion function $\mathcal{T}(x,y)$ is calibrated with the reference stars
and applied to the measured position of the target star to estimate
its real position the sky . Then the movement of the target star
  is measured relative to the barycenter of the reference stars. This
procedure must be repeated for each exposure because the distortion
function can change due to the thermal expansion for instance.

\begin{figure}
    \centering
    \includegraphics[width=0.6\hsize]{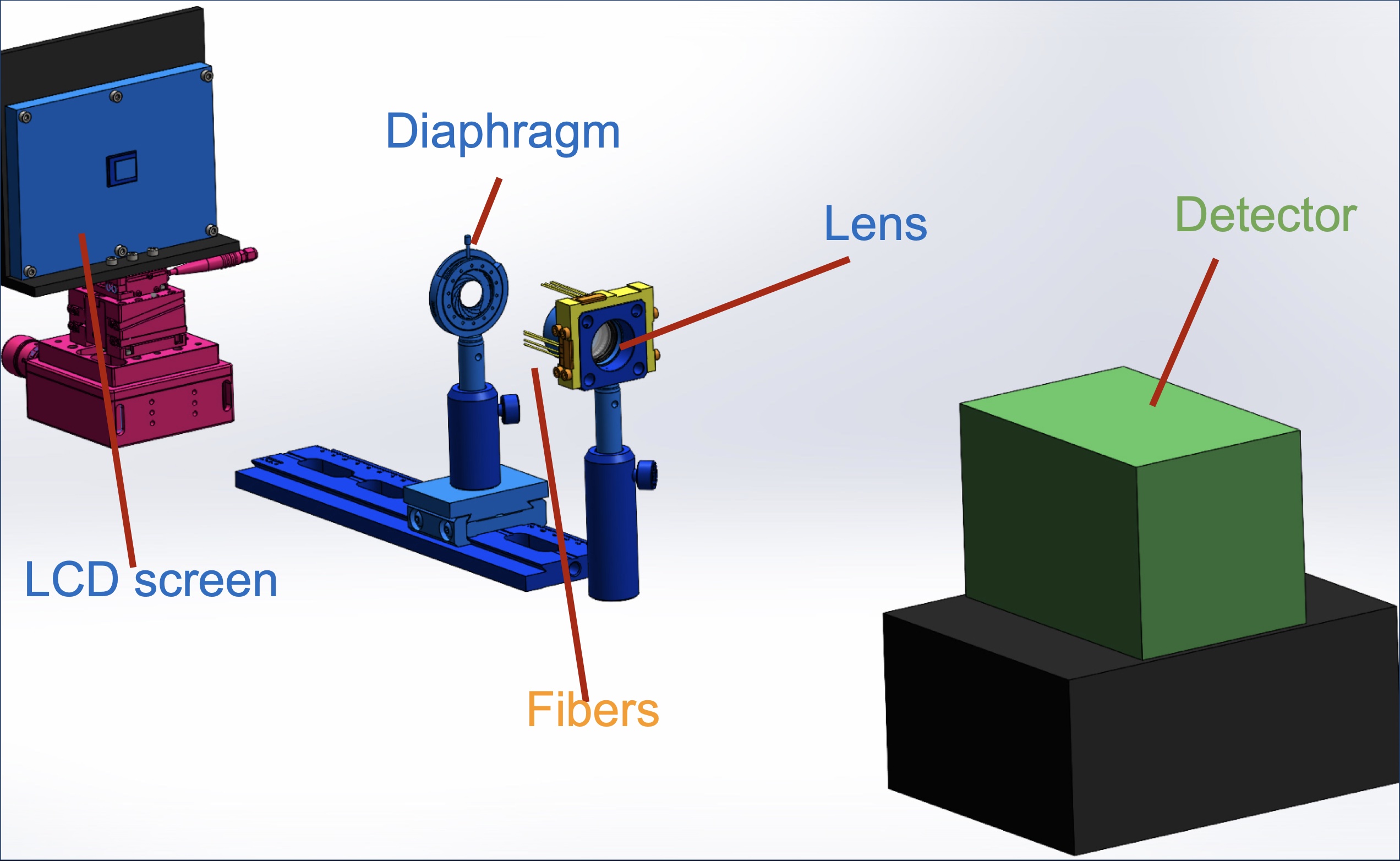}
    \caption{Testbed for distortion calibration. The sources on the
      LCD screen are images onto the detector in a $2f-2f$ optical
      setup with no magnification factor. By moving the
      position of the diaphragm, one will change the distortion of the
    image. Using the reference stars, we will retriev the distortion function.}
    \label{fig:distortion-calib}
  \end{figure}

We plan now to reproduce the simulation in the lab
(Fig.~\ref{fig:distortion-calib}) using a simple optical bench where
we change the distortion by moving the pupil position in the $z$-direction.

\section{Conclusion}

Theia is a project for an astrometric observatory based on high-precision
differential astrometry measurements on a limited single field
(0.5\,deg$\times$0.5\,deg). The two main science cases are nature of
dark matter and Earth-like planets around solar neighbourhood. Theia’s
payload is a 0.8\,m diameter diffraction limited TMA Korsch
telescope. New very large format CMOS visible detectors are being
investigated in order to cope with a diffraction-limited yet
large field of view without too many detectors. Telescope stability impacts
performances except if one is able to calibrate the telescope
distortion during each exposure duration (e.g.\ 0.1\,s) using reference stars with \gaia positions.

\acknowledgements{ The authors would like to acknowledge the
  contributions of the researchers and engineers who are not
  co-authors of this article but who have participated in the proposed
  missions and provided valuable input in response to ESA's successive
  calls for proposals: NEAT (M3), micro-NEAT (S1), and Theia (M4,
  M5, M7).

  The co-authors would like also to express their gratitude to Alberto Riva
  for presenting this contribution at the conference session.

  With regard to the funding of our research, we would like to
  acknowledge the support of the LabEx FOCUS ANR-11-LABX-0013 and the
  CNES agency. ML would like to acknowledge the support of her PhD
  grant from CNES and Pyxalis.
  
  This research has made use of NASA’s Astrophysics Data System
  Bibliographic Services and of the Aladin sky atlas, CDS, Strasbourg
  Astronomical Observatory, France.

\bibliography{theia-spie-2024} 

\begin{thebibliography}{10}

\bibitem{1980CeMec..22..153K}
{Kovalevsky}, J., ``{Global Astrometry by Space Techniques},'' {\em Celestial
  Mechanics}~{\bf 22},  153--163 (Aug. 1980).

\bibitem{1997ESASP1200.....E}
{ESA}, ed.,  [{\em {The HIPPARCOS and TYCHO catalogues. Astrometric and
  photometric star catalogues derived from the ESA HIPPARCOS Space Astrometry
  Mission}}{\nolinebreak\hspace{0.1em}]}, {\em ESA Special Publication} {\bf
  1200} (1997).

\bibitem{2016A&A...595A...1G}
{Gaia Collaboration}, ``{The Gaia mission},'' {\em \aap}~{\bf 595},  A1 (Nov.
  2016).

\bibitem{Malbet+12}
{Malbet}, F., {L{\'e}ger}, A., {Shao}, M., {Goullioud}, R., {Lagage}, P.-O.,
  {Brown}, A. G.~A., {Cara}, C., {Durand }, G., {Eiroa}, C., {Feautrier}, P.,
  {Jakobsson}, B., {Hinglais}, E., {Kaltenegger}, L., {Labadie}, L.,
  {Lagrange}, A.-M., {Laskar}, J., {Liseau}, R., {Lunine}, J., {Maldonado}, J.,
  {Mercier}, M., {Mordasini}, C., {Queloz}, D., {Quirrenbach}, A., {Sozzetti},
  A.~r., {Traub}, W., {Absil}, O., {Alibert}, Y., {Andrei}, A.~H., {Arenou},
  F., {Beichman}, C., {Chelli}, A., {Cockell}, C.~S., {Duvert}, G.,
  {Forveille}, T., {Garcia}, P. J.~V., {Hobbs}, D., {Krone-Martins}, A.,
  {Lammer}, H., {Meunier}, N., {Minardi}, S., {Moitinho de Almeida}, A.,
  {Rambaux}, N., {Raymond}, S., {R{\"o}ttgering}, H. J.~A., {Sahlmann}, J.,
  {Schuller}, P.~A., {S{\'e}gransan}, D., {Selsis}, F., {Surdej}, J.,
  {Villaver}, E., {White}, G.~J., and {Zinnecker}, H., ``High precision
  astrometry mission for the detection and characterization of nearby habitable
  planetary systems with the nearby earth astrometric telescope (neat),'' {\em
  Exp. Ast.}~{\bf 34},  385--413 (Oct. 2012).

\bibitem{2017arXiv170701348T}
{The Theia Collaboration}, {Boehm}, C., and {et al.}, ``{Theia: Faint objects
  in motion or the new astrometry frontier},'' {\em arXiv e-prints} ,
  arXiv:1707.01348 (July 2017).

\bibitem{2022SPIE12180E..1FM}
{Malbet}, F., {Labadie}, L., {Sozzetti}, A., {Mamon}, G.~A., {Shao}, M.,
  {Goullioud}, R., {L{\'e}ger}, A., {Gai}, M., {Riva}, A., {Busonero}, D.,
  {L{\'e}pine}, T., {Lizzana}, M., {Brandeker}, A., and {Villaver}, E.,
  ``{Theia: science cases and mission profiles for high precision astrometry in
  the future},'' in [{\em Space Telescopes and Instrumentation 2022: Optical,
  Infrared, and Millimeter Wave}{\nolinebreak\hspace{0.1em}]},  {Coyle}, L.~E.,
  {Matsuura}, S., and {Perrin}, M.~D., eds., {\em Society of Photo-Optical
  Instrumentation Engineers (SPIE) Conference Series} {\bf 12180},  121801F
  (Aug. 2022).

\bibitem{Malbet+21}
{Malbet}, F., {Boehm}, C., {Krone-Martins}, A., {Amorim}, A.,
  {Anglada-Escud{\'e}}, G., {Brand eker}, A., {Courbin}, F., {En{\ss}lin}, T.,
  {Falc{\~a}o}, A., {Freese}, K., {Holl}, B., {Labadie}, L., {L{\'e}ger}, A.,
  {Mamon}, G.~A., {McArthur}, B., {Mora}, A., {Shao}, M., {Sozzetti}, A.~r.,
  {Spolyar}, D., {Villaver}, E., {Abbas}, U., {Albertus}, C., {Alves}, J.,
  {Barnes}, R., {Bonomo}, A.~S., {Bouy}, H., {Brown}, W.~R., {Cardoso}, V.,
  {Castellani}, M., {Chemin}, L., {Clark}, H., {Correia}, A. r. C.~M.,
  {Crosta}, M., {Crouzier}, A., {Damasso}, M., {Darling}, J., {Davies}, M.~B.,
  {Diaferio}, A., {Fortin}, M., {Fridlund}, M., {Gai}, M., {Garcia}, P.,
  {Gnedin}, O., {Goobar}, A., {Gordo}, P., {Goullioud}, R., {Hall}, D.,
  {Hambly}, N., {Harrison}, D., {Hobbs}, D., {Holland }, A., {H{\o}g}, E.,
  {Jordi}, C., {Klioner}, S., {Lan{\c{c}}on}, A., {Laskar}, J., {Lattanzi}, M.,
  {Le Poncin-Lafitte}, C., {Luri}, X., {Michalik}, D., {de Almeida}, A.~M.,
  {Mour{\~a}o}, A., {Moustakas}, L., {Murray}, N.~J., {Muterspaugh}, M.,
  {Oertel}, M., {Ostorero}, L., {Portell}, J., {Prost}, J.-P., {Quirrenbach},
  A., {Schneider}, J., {Scott}, P., {Siebert}, A., {Silva}, A.~d., {Silva}, M.,
  {Th{\'e}bault}, P., {Tomsick}, J., {Traub}, W., {de Val-Borro}, M.,
  {Valluri}, M., {Walton}, N.~A., {Watkins}, L.~L., {White}, G., {Wyrzykowski},
  L., {Wyse}, R., and {Yamada}, Y., ``Faint objects in motion: the new frontier
  of high precision astrometry,'' {\em Exp. Ast.}~{\bf 51},  845--886 (June
  2021).

\bibitem{2021pdaa.book.....N}
{National Academies of Sciences}, E. and Medicine,  [{\em {Pathways to
  Discovery in Astronomy and Astrophysics for the
  2020s}}{\nolinebreak\hspace{0.1em}]} (2021).

\bibitem{Meunier&Lagrange22}
{Meunier}, N. and {Lagrange}, A.~M., ``{A new estimation of astrometric
  exoplanet detection limits in the habitable zone around nearby stars},'' {\em
  \aap}~{\bf 659},  A104 (Mar. 2022).

\bibitem{Bryson+21}
{Bryson}, S., {Kunimoto}, M., {Kopparapu}, R.~K., {Coughlin}, J.~L., {Borucki},
  W.~J., {Koch}, D., {Aguirre}, V.~S., {Allen}, C., {Barentsen}, G., {Batalha},
  N.~M., {Berger}, T., {Boss}, A., {Buchhave}, L.~A., {Burke}, C.~J.,
  {Caldwell}, D.~A., {Campbell}, J.~R., {Catanzarite}, J., {Chand rasekaran},
  H., {Chaplin}, W.~J., {Christiansen}, J.~L., {Christensen-Dalsgaard}, J.,
  {Ciardi}, D.~R., {Clarke}, B.~D., {Cochran}, W.~D., {Dotson}, J.~L., {Doyle},
  L.~R., {Duarte}, E.~S., {Dunham}, E.~W., {Dupree}, A.~K., {Endl}, M.,
  {Fanson}, J.~L., {Ford}, E.~B., {Fujieh}, M., {Gautier}, Thomas~N., I.,
  {Geary}, J.~C., {Gilliland }, R.~L., {Girouard}, F.~R., {Gould}, A., {Haas},
  M.~R., {Henze}, C.~E., {Holman}, M.~J., {Howard}, A.~W., {Howell}, S.~B.,
  {Huber}, D., {Hunter}, R.~C., {Jenkins}, J.~M., {Kjeldsen}, H.,
  {Kolodziejczak}, J., {Larson}, K., {Latham}, D.~W., {Li}, J., {Mathur}, S.,
  {Meibom}, S., {Middour}, C., {Morris}, R.~L., {Morton}, T.~D., {Mullally},
  F., {Mullally}, S.~E., {Pletcher}, D., {Prsa}, A., {Quinn}, S.~N.,
  {Quintana}, E.~V., {Ragozzine}, D., {Ramirez}, S.~V., {Sand erfer}, D.~T.,
  {Sasselov}, D., {Seader}, S.~E., {Shabram}, M., {Shporer}, A., {Smith},
  J.~C., {Steffen}, J.~H., {Still}, M., {Torres}, G., {Troeltzsch}, J.,
  {Twicken}, J.~D., {Uddin}, A.~K., {Van Cleve}, J.~E., {Voss}, J., {Weiss},
  L.~M., {Welsh}, W.~F., {Wohler}, B., and {Zamudio}, K.~A., ``The occurrence
  of rocky habitable-zone planets around solar-like stars from kepler data,''
  {\em \aj}~{\bf 161},  36 (Jan. 2021).

\bibitem{Vitral+22}
{Vitral}, E., {Kremer}, K., {Libralato}, M., {Mamon}, G.~A., and {Bellini}, A.,
  ``{Stellar graveyards: clustering of compact objects in globular clusters NGC
  3201 and NGC 6397},'' {\em \mnras}~{\bf 514},  806--825 (July 2022).

\bibitem{Lindegren+2018}
{Lindegren}, L., {Hern{\'a}ndez}, J., {Bombrun}, A., {Klioner}, S., {Bastian},
  U., {Ramos-Lerate}, M., {de Torres}, A., {Steidelm{\"u}ller}, H.,
  {Stephenson}, C., {Hobbs}, D., {Lammers}, U., {Biermann}, M., {Geyer}, R.,
  {Hilger}, T., {Michalik}, D., {Stampa}, U., {McMillan}, P.~J.,
  {Casta{\~n}eda}, J., {Clotet}, M., {Comoretto}, G., {Davidson}, M.,
  {Fabricius}, C., {Gracia}, G., {Hambly}, N.~C., {Hutton}, A., {Mora}, A.,
  {Portell}, J., {van Leeuwen}, F., {Abbas}, U., {Abreu}, A., {Altmann}, M.,
  {Andrei}, A., {Anglada}, E., {Balaguer-N{\'u}{\~n}ez}, L., {Barache}, C.,
  {Becciani}, U., {Bertone}, S., {Bianchi}, L., {Bouquillon}, S., {Bourda}, G.,
  {Br{\"u}semeister}, T., {Bucciarelli}, B., {Busonero}, D., {Buzzi}, R.,
  {Cancelliere}, R., {Carlucci}, T., {Charlot}, P., {Cheek}, N., {Crosta}, M.,
  {Crowley}, C., {de Bruijne}, J., {de Felice}, F., {Drimmel}, R., {Esquej},
  P., {Fienga}, A., {Fraile}, E., {Gai}, M., {Garralda}, N.,
  {Gonz{\'a}lez-Vidal}, J.~J., {Guerra}, R., {Hauser}, M., {Hofmann}, W.,
  {Holl}, B., {Jordan}, S., {Lattanzi}, M.~G., {Lenhardt}, H., {Liao}, S.,
  {Licata}, E., {Lister}, T., {L{\"o}ffler}, W., {Marchant}, J.,
  {Martin-Fleitas}, J.~M., {Messineo}, R., {Mignard}, F., {Morbidelli}, R.,
  {Poggio}, E., {Riva}, A., {Rowell}, N., {Salguero}, E., {Sarasso}, M.,
  {Sciacca}, E., {Siddiqui}, H., {Smart}, R.~L., {Spagna}, A., {Steele}, I.,
  {Taris}, F., {Torra}, J., {van Elteren}, A., {van Reeven}, W., and
  {Vecchiato}, A., ``{Gaia Data Release 2. The astrometric solution},'' {\em
  \aap}~{\bf 616},  A2 (Aug. 2018).

\bibitem{sanderson+2019}
{WFIRST Astrometry Working Group}, {Sanderson}, R.~E., {Bellini}, A.,
  {Casertano}, S., {Lu}, J.~R., {Melchior}, P., {Libralato}, M., {Bennett}, D.,
  {Shao}, M., {Rhodes}, J., {Sohn}, S.~T., {Malhotra}, S., {Gaudi}, S., {Fall},
  S.~M., {Nelan}, E., {Guhathakurta}, P., {Anderson}, J., and {Ho}, S.,
  ``{Astrometry with the Wide-Field Infrared Space Telescope},'' {\em J. of
  Astr. Teles., Instr., and Syst.}~{\bf 5},  044005 (Oct. 2019).

\bibitem{Gravity+2021}
{Gravity Collaboration}, {Abuter}, R., {Amorim}, A., {Baub{\"o}ck}, M.,
  {Berger}, J.~P., {Bonnet}, H., {Brandner}, W., {Cl{\'e}net}, Y., {Davies},
  R., {de Zeeuw}, P.~T., {Dexter}, J., {Dallilar}, Y., {Drescher}, A.,
  {Eckart}, A., {Eisenhauer}, F., {F{\"o}rster Schreiber}, N.~M., {Garcia}, P.,
  {Gao}, F., {Gendron}, E., {Genzel}, R., {Gillessen}, S., {Habibi}, M.,
  {Haubois}, X., {Hei{\ss}el}, G., {Henning}, T., {Hippler}, S., {Horrobin},
  M., {Jim{\'e}nez-Rosales}, A., {Jochum}, L., {Jocou}, L., {Kaufer}, A.,
  {Kervella}, P., {Lacour}, S., {Lapeyr{\`e}re}, V., {Le Bouquin}, J.~B.,
  {L{\'e}na}, P., {Lutz}, D., {Nowak}, M., {Ott}, T., {Paumard}, T., {Perraut},
  K., {Perrin}, G., {Pfuhl}, O., {Rabien}, S., {Rodr{\'\i}guez-Coira}, G.,
  {Shangguan}, J., {Shimizu}, T., {Scheithauer}, S., {Stadler}, J., {Straub},
  O., {Straubmeier}, C., {Sturm}, E., {Tacconi}, L.~J., {Vincent}, F., {von
  Fellenberg}, S., {Waisberg}, I., {Widmann}, F., {Wieprecht}, E., {Wiezorrek},
  E., {Woillez}, J., {Yazici}, S., {Young}, A., and {Zins}, G., ``{Improved
  GRAVITY astrometric accuracy from modeling optical aberrations},'' {\em
  \aap}~{\bf 647},  A59 (Mar. 2021).

\bibitem{Crouzier+16}
{Crouzier}, A., {Malbet}, F., {Henault}, F., {L{\'e}ger}, A., {Cara}, C.,
  {LeDuigou}, J.~M., {Preis}, O., {Kern}, P., {Delboulbe}, A., {Martin}, G.,
  {Feautrier}, P., {Stadler}, E., {Lafrasse}, S., {Rochat}, S., {Ketchazo}, C.,
  {Donati}, M., {Doumayrou}, E., {Lagage}, P.~O., {Shao}, M., {Goullioud}, R.,
  {Nemati}, B., {Zhai}, C., {Behar}, E., {Potin}, S., {Saint-Pe}, M., and
  {Dupont}, J., ``A detector interferometric calibration experiment for high
  precision astrometry,'' {\em \aap}~{\bf 595},  A108 (Nov. 2016).

\end{thebibliography}
\bibliographystyle{spiebib} 

\end{document}